\documentclass[prb,twocolumn,showpacs,superscriptaddress,preprintnumbers]{revtex4}
\usepackage{amsmath}
\usepackage{graphicx}
\usepackage{dcolumn}
\usepackage{bm}

\begin{document}

\title{Correlated electron transport through parallel double-quantum-dot}
\author{Rong L\"{u}$^{1}$, Zhi-Rong Liu$^{2}$, and Guang-Ming Zhang$^{2,}$}
\affiliation{Center for Advanced Study, Tsinghua University, Beijing 100084, China; \\
$^{2}$Department of Physics, Tsinghua University, Beijing 100084, China}
\date{\today}

\begin{abstract}
We investigate the spectral and transport properties of parallel
double-quantum-dot (DQD) system with interdot tunneling coupling
in both the equilibrium and nonequilibrium cases. The special
geometry of DQD system is considered, in which each dot is
connected to two leads by the tunneling barriers. With the help of
Keldysh nonequilibrium Green function technique and the
equation-of-motion approach, the spectral function and the
conductance spectra of DQD system are calculated in two cases with
and without the intradot Coulomb interaction, respectively. The
exact calculation is performed in the absence of intradot Coulomb
interaction. For the case with intradot Coulomb interaction, the
Hartree-Fock approximation is applied to truncate the equation of
motion for the high-order Green functions at high temperatures.
The phenomenon of correlated electron transport is clearly shown
in the linear conductance of each dot in the presence of interdot
tunneling when setting one dot level and tuning another. The
interplay between the intradot Coulomb interaction and the
interdot tunneling coupling is displayed.
\end{abstract}

\maketitle

\section{Introduction}

Recently, due to the rapid progress in nanotechnologies, quantum
transport through DQD system has been the subject of active
theoretical and experimental research.\cite{van der Wiel 03}
Compared to the single quantum dot, the interdot coupling and the
intradot on-site Coulomb interaction in DQD system could generate
novel many-body states, e. g. the molecular Kondo
state.\cite{experiments} Apart from its importance in
understanding some basic problems in condensed matter physics, the
DQD devices are also crucial ingredients in the emerging field of
spintronics and several quantum computation schemes designed with
the electron spins or with the coherent mode in an artificial
molecule.\cite{Wolf 01,Awschalom 02}

Transport through DQD system in the Coulomb blockade and the Kondo regimes
has already received some theoretical attention, in which two dots are
arranged in series,\cite{Ruzin 92,Fong92,Golden
96,Datta,Middleton93,Stafford94,Matveev96,Georges99,Aguado00,Aono01,Izumida00,Lopez02}
parallel\cite{Kikoin01,Lopez02,Zhang04} or T-shape.\cite%
{Kim01,Boese02,Cornaglia04} In most of these geometries, the
conductance spectra have been studied for electron transport
through the whole DQD system, while the electron transport through
each individual dot could not be detected. In this paper we
consider a new configuration of DQD system (see Fig. \ref{fig01}),
in which each quantum dot is connected to two leads by the
tunneling barriers. In this geometry, the collected electron
transport phenomenon can be investigated by measuring the
transport properties through one individual dot when tuning the
level of another dot and its interdot coupling. The
similar configuration of DQD system has been proposed in Ref. %
\onlinecite{Wilhelm 01}, however, the detailed investigation on
the spectral and conductance spectra, and the interplay between
interdot tunneling coupling and intradot on-site Coulomb
interaction has been not discussed yet.

In this paper, by using the Keldysh nonequilibrium Green function
technique and equation-of-motion method, we calculate the spectral
and the conductance spectra of tunneling-coupled DQD with and
without intradot Coulomb interaction, respectively, in both the
equilibrium limit and in the nonequilibrium case. In Sec. II, we
formulate the model Hamiltonian of the special geometry of DQD
system, and deduce the current formula through each quantum dot.

In Sec. III, for DQD without the intradot Coulomb interaction, we
exactly calculate the Green functions, from which both the
spectral function and the conductance are obtained, showing the
resonant peaks corresponding to the molecular resonant-tunneling
bonding and anti-bonding states. The image of total differential
conductance shows ''anticrossing'' phenomenon when the levels of
two dots match. It is interesting to find that in the presence of
interdot tunneling coupling, the electron transport through one
dot is strongly influenced by the electron transport through
another dot, namely, when setting dot 1 level ($\epsilon _{1}$)
and tuning dot 2 level ($\epsilon _{2}$) the linear conductance of
dot 1 ($G_{dot1}$) shows a dip at the same
position of $\epsilon _{2}$, where the linear conductance of dot 2 ($%
G_{dot2} $) displays a resonant peak. Increasing the interdot tunneling
coupling gives rise to a more complicate structure of linear conductances
through each dot.

In Sec. IV, we apply the Hartree-Fock approximation to DQD with the intradot
Coulomb interaction to truncate the equation of motion for the high-order
Green functions, which is known to capture the correct qualitative feature
of physics of single quantum dot in the Coulomb blockade regime. Under the
same approximation, we also obtain the Keldysh less Green function, which is
needed in the self-consistent evaluation of the dot occupation numbers when
a finite bias voltage is applied across the system. As the intradot Coulomb
interaction becomes the largest energy scale in the single-electron
tunneling regime of quantum dot, the intradot Coulomb interaction leads to
two groups of peaks in both the spectral and conductance spectra, and the
interdot tunneling coupling causes the splitting of peaks in each group.
Compared to the case without the intradot Coulomb interaction, when setting $%
\epsilon _{1}$ and tuning $\epsilon _{2}$, one more dip in
$G_{dot1}$ and resonant peak in $G_{dot2}$ will appear
corresponding to the Coulomb interaction energy, which also
provides the clear evidence of the correlated electron transport
through DQD system. The paper is closed with a brief summary in
Sec. V.

\section{Physical Model And Current Formula}

\begin{figure}[tbp]
\includegraphics[width=7cm]{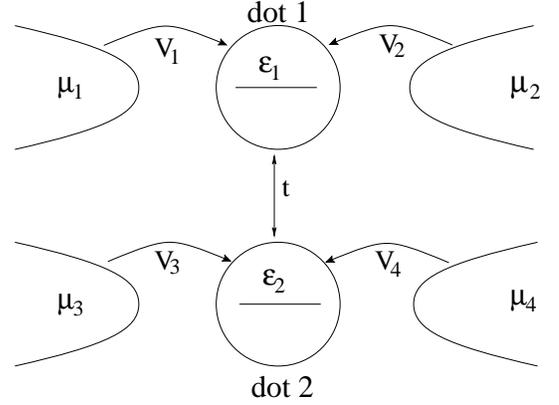}
\caption{Schematic diagram of the parallel DQD system studied in this paper.}
\label{fig01}
\end{figure}

Quantum dots behave as artificial atoms and single-electron
transistors in their charge and energy quantizations due to the
small dimensions (compared to the Fermi wavelength). They are
often described by the single-impurity Anderson model, in which
there is a Coulomb repulsion between electrons in the dot. The
parallel DQD system could be modelled by using a two-impurity
Anderson model with an extra interdot tunneling corresponding to
the electrons hopping between two dots. Each QD is attached to two
electronic leads with different chemical potentials by
quantum-tunneling barriers. Geometry of the system is shown
schematically in Fig. \ref{fig01}. The model Hamiltonian is then
given by
\begin{equation}
H=H_{leads}+H_{DD}+H_{T}.  \label{Hamiltonian}
\end{equation}%
The first term $H_{leads}$ describes the noninteracting electrons in the $%
\eta $-th lead with the electron creation $c_{\mathbf{k\eta }\sigma
}^{\dagger }$ and annihilation $c_{\mathbf{k\eta }\sigma }$ operators,
\begin{equation}
H_{leads}=\sum_{\mathbf{k},\eta ,\sigma }\epsilon _{\mathbf{k\eta }}c_{%
\mathbf{k\eta }\sigma }^{\dagger }c_{\mathbf{k\eta }\sigma },
\end{equation}%
where $\eta =1$, $2$, $3$, $4$ corresponding to the four leads shown in Fig. %
\ref{fig01}, $\mathbf{k}$ denotes the wave vector, and $\sigma $
is the spin index. The second term $H_{DD}$ in Eq.
(\ref{Hamiltonian}) describes the electrons in the dots 1 and 2
with the interdot tunneling coupling $t$ and
the intradot on-site Coulomb interaction $U$,%
\begin{eqnarray}
H_{DD} &=&\sum_{\sigma }\epsilon _{1}d_{1\sigma }^{\dagger }d_{1\sigma }+%
\frac{U}{2}\sum_{\sigma }n_{1\sigma }n_{1\overline{\sigma }}  \notag \\
&&+\sum_{\sigma }\epsilon _{2}d_{2\sigma }^{\dagger }d_{2\sigma }+\frac{U}{2}%
\sum_{\sigma }n_{2\sigma }n_{2\overline{\sigma }}  \notag \\
&&-t\sum_{\sigma }\left( d_{1\sigma }^{\dagger }d_{2\sigma }+d_{2\sigma
}^{\dagger }d_{1\sigma }\right) .
\end{eqnarray}%
The last term in Eq. (\ref{Hamiltonian}) describes the tunneling coupling
between dots and the external leads with the tunneling matrix elements $V_{\eta}$,%
\begin{eqnarray}
H_{T} &=&\sum_{\mathbf{k},\sigma }\left( V_{1}c_{\mathbf{k}1\sigma
}^{\dagger }d_{1\sigma }+V_{2}c_{\mathbf{k}2\sigma }^{\dagger }d_{1\sigma
}+h.c.\right)   \notag \\
&&+\sum_{\mathbf{k},\sigma }\left( V_{3}c_{\mathbf{k}3\sigma }^{\dagger
}d_{2\sigma }+V_{4}c_{\mathbf{k}4\sigma }^{\dagger }d_{2\sigma }+h.c.\right)
.
\end{eqnarray}%

When we use the Keldysh nonequilibrium Green function technique to
express the current through dots 1 and 2, the respective
conductance can be written in terms of the distribution functions
of leads and the local properties of DQD system. To this end we
write the current from the lead 1 to dot 1 as
\begin{equation}
J_{1}=-e\left\langle dN_{1}/dt\right\rangle =-\left( ie/\hbar \right)
\left\langle \left[ H,N_{1}\right] \right\rangle ,
\end{equation}%
where $N_{1}=\sum_{\mathbf{k},\sigma }c_{\mathbf{k}1\sigma }^{\dagger }c_{%
\mathbf{k}1\sigma }$. With the help of Eq. (\ref{Hamiltonian}), one can
easily obtain the current from the lead 1 to dot 1 as%
\begin{eqnarray}
J_{1} &=&\frac{ie}{\hbar }\sum_{\mathbf{k},\sigma }\left( V_{1}\left\langle
c_{\mathbf{k}1\sigma }^{\dagger }d_{1\sigma }\right\rangle -V_{1}^{\ast
}\left\langle d_{1\sigma }^{\dagger }c_{\mathbf{k}1\sigma }\right\rangle
\right)  \notag \\
&=&\frac{e}{\hbar }\sum_{\mathbf{k},\sigma }\int \frac{d\omega }{2\pi }%
\left[ V_{1}G_{1\sigma ,\mathbf{k}1\sigma }^{<}\left( \omega
\right) -V_{1}^{\ast }G_{\mathbf{k}1\sigma ,1\sigma }^{<}\left(
\omega \right) \right],
\end{eqnarray}%
where the Keldysh less Green functions are defined as
\begin{eqnarray}
G_{1\sigma ,\mathbf{k}1\sigma }^{<}\left( t,t^{\prime }\right) &\equiv
&i\left\langle c_{\mathbf{k}1\sigma }^{\dagger }\left( t^{\prime }\right)
d_{1\sigma }\left( t\right) \right\rangle ,  \notag \\
G_{\mathbf{k}1\sigma ,1\sigma }^{<}\left( t,t^{\prime }\right) &\equiv
&i\left\langle d_{1\sigma }^{\dagger }\left( t^{\prime }\right) c_{\mathbf{k}%
1\sigma }\left( t\right) \right\rangle .
\end{eqnarray}

By using the Keldysh Green function formalism, one can obtain the Dyson
equations as%
\begin{eqnarray}
G_{\mathbf{k}1\sigma ,1\sigma }^{<}\left( \omega \right)  &=&V_{1}\left[ g_{%
\mathbf{k}1\sigma }^{t}\left( \omega \right) G_{1\sigma ,1\sigma }^{<}\left(
\omega \right) \right.   \notag \\
&&\left. -g_{\mathbf{k}1\sigma }^{<}\left( \omega \right) G_{1\sigma
,1\sigma }^{\overline{t}}\left( \omega \right) \right] ,  \notag \\
G_{1\sigma ,\mathbf{k}1\sigma }^{<}\left( \omega \right)  &=&V_{1}^{\ast }%
\left[ g_{\mathbf{k}1\sigma }^{<}\left( \omega \right) G_{1\sigma ,1\sigma
}^{t}\left( \omega \right) \right.   \notag \\
&&\left. -g_{\mathbf{k}1\sigma }^{\overline{t}}\left( \omega \right)
G_{1\sigma ,1\sigma }^{<}\left( \omega \right) \right] ,
\end{eqnarray}%
where the Keldysh less Green function for electrons in dot $1$ is
defined as
\begin{equation}
G_{1\sigma ,1\sigma }^{<}\left( t,t^{\prime }\right) \equiv
i\left\langle d_{1\sigma }^{\dagger }\left( t^{\prime }\right)
d_{1\sigma }\left( t\right) \right\rangle ,
\end{equation}
and $g_{\mathbf{k}1\sigma }$ is the Green functions for
noninteracting electrons of the lead 1. The time-ordered and
anti-time-ordered Green functions are denoted by the superscripts $t$ and $%
\overline{t}$, respectively. Therefore, the current from the lead 1 (or 2)
to the dot 1 can be expressed as%
\begin{eqnarray}
J_{1(2)} &=&\frac{ie}{\hbar }\sum_{\sigma }\int \frac{d\epsilon }{2\pi }%
\Gamma _{1(2)}\left( \epsilon \right) \left\{ G_{1\sigma ,1\sigma
}^{<}\left( \epsilon \right) \right.   \notag \\
&&\left. +f_{1(2)}\left( \epsilon \right) \left[ G_{1\sigma ,1\sigma
}^{r}\left( \epsilon \right) -G_{1\sigma ,1\sigma }^{a}\left( \epsilon
\right) \right] \right\} ,
\end{eqnarray}%
where the line-width $\Gamma _{1(2)}=2\pi \rho _{F}V_{1(2)}^{2}$
with $\rho _{F}$ the density of states of leads, and
$f_{1(2)}\left( \epsilon \right) $ is the Fermi-Dirac distribution
functions of the lead 1 (or 2) with the chemical potential $\mu
_{1(2)}$. Here we have assumed that the leads give rise to a flat,
energy independent, density of states (i. e., the wide-band
limit). By applying the similar procedure, the currents from the
leads 3 and 4 to the dot 2 are found
similarly,%
\begin{eqnarray}
J_{3(4)} &=&\frac{ie}{\hbar }\sum_{\sigma }\int \frac{d\epsilon }{2\pi }%
\Gamma _{3(4)}\left( \epsilon \right) \left\{ G_{2\sigma ,2\sigma
}^{<}\left( \epsilon \right) \right.   \notag \\
&&\left. +f_{3(4)}\left( \epsilon \right) \left[ G_{2\sigma ,2\sigma
}^{r}\left( \epsilon \right) -G_{2\sigma ,2\sigma }^{a}\left( \epsilon
\right) \right] \right\} ,
\end{eqnarray}%
with $\Gamma _{3(4)}=2\pi \rho _{F}V_{3(4)}^{2}$ for the lead 3 (or 4).

Under the symmetric condition: $\mu _{1}+\mu _{2}=\mu _{3}+\mu _{4}$, there
is no current between dot 1 and dot 2. In the steady state, the current will
be uniform, so that the current through dot 1 satisfies $%
J_{dot1}=J_{1}=-J_{2}$, and then one can symmetrize the current through the
dot 1 as%
\begin{eqnarray}
J_{dot1} &=&\frac{1}{2}\left( J_{1}-J_{2}\right)   \notag \\
&=&i\frac{e}{2\hbar }\sum_{\sigma }\int \frac{d\epsilon }{2\pi
}\left[\left( f_{1}\Gamma _{1}-f_{2}\Gamma _{2}\right) \left(
G_{1\sigma ,1\sigma
}^{r}\left( \epsilon \right) \right.\right.   \notag \\
&&\left.\left. -G_{1\sigma ,1\sigma }^{a}\left( \epsilon \right)
\right) +\left( \Gamma _{1}-\Gamma _{2}\right) G_{1\sigma ,1\sigma
}^{<}\left( \epsilon \right)\right] ,  \label{current1}
\end{eqnarray}%
and similarly the current through the dot 2 is%
\begin{eqnarray}
J_{dot2} &=&i\frac{e}{2\hbar }\sum_{\sigma }\int \frac{d\epsilon }{2\pi }\left[%
\left( f_{3}\Gamma _{3}-f_{4}\Gamma _{4}\right) \left( G_{2\sigma
,2\sigma
}^{r}\left( \epsilon \right) \right.\right.   \notag \\
&&\left.\left. -G_{2\sigma ,2\sigma }^{a}\left( \epsilon \right)
\right) +\left( \Gamma _{3}-\Gamma _{4}\right) G_{2\sigma ,2\sigma
}^{<}\left( \epsilon \right)\right] .  \label{current2}
\end{eqnarray}%
It is noted that Eqs. (\ref{current1}) and (\ref{current2}) give
rise to the current through each dot of the DQD system in terms of
the distribution functions of leads and the local properties of
dots. In order to obtain the current, one has to compute the
retarded and the Keldysh less Green functions of the DQD system in
the presence of both the interdot tunneling coupling $t$ and
intradot Coulomb interaction $U$ as well as the tunneling coupling
of DQD system into the leads.

Without the loss of generality, we assume that the dot-lead
tunneling coupling are symmetric: $\Gamma _{1}=\Gamma _{2}=\Gamma
_{3}=\Gamma _{4}=\Gamma $, and the symmetric configuration of the
chemical potentials: $\mu _{1}=\mu _{3}=\mu +eV/2$, and $\mu
_{2}=\mu _{4}=\mu -eV/2$. Then the current through the dot
1 (or 2) becomes%
\begin{equation}
J_{dot1(2)}=\frac{e\Gamma }{2\hbar }\sum_{\sigma }\int d\epsilon \left[
f_{1}\left( \epsilon \right) -f_{2}\left( \epsilon \right) \right] \mathcal{A%
}_{1\left( 2\right) \sigma }\left( \epsilon \right) ,
\end{equation}%
where the dot spectral functions are defined as
\begin{equation}
\mathcal{A}_{1\left( 2\right) \sigma }=\left( -1/\pi \right) Im\left[
G_{1\left( 2\right) \sigma ,1\left( 2\right) \sigma }^{r}\right] .
\end{equation}
The associated differential conductance at $\mu =0$ is found to be%
\begin{eqnarray}
&&\frac{dJ_{dot1(2)}}{dV} =\frac{e^{2}\beta \Gamma }{4\hbar
}\sum_{\sigma }\int d\epsilon \mathcal{A}_{1\left( 2\right) \sigma
}\left( \epsilon \right)
\notag \\
&&\times \left[ \frac{e^{\beta \left( \epsilon
-\frac{eV}{2}\right) }}{\left(
e^{\beta \left( \epsilon -\frac{eV}{2}\right) }+1\right) ^{2}}+\frac{%
e^{\beta \left( \epsilon +\frac{eV}{2}\right) }}{\left( e^{\beta \left(
\epsilon +\frac{eV}{2}\right) }+1\right) ^{2}}\right] ,
\label{differential-conductance}
\end{eqnarray}%
and the linear conductance at zero bias is thus given by%
\begin{equation}
G_{dot1(2)}=\frac{e^{2}\Gamma }{2\hbar }\sum_{\sigma }\int d\epsilon \left( -%
\frac{\partial f_{FD}\left( \epsilon \right) }{\partial \epsilon }\right)
\mathcal{A}_{1\left( 2\right) \sigma }\left( \epsilon \right) ,
\label{linear-conductance}
\end{equation}%
where $f_{FD}\left(\epsilon\right)$ is the Fermi-Dirac
distribution function.

\section{The noninteracting dots}

In this section we study the case of noninteracting dots, i. e., $U=0$. In
this case, we can exactly derive the Green functions, the associated
spectral, and transport properties of the DQD system, which shows the clear
evidence of correlated electron transport through DQD.

By applying the equation-of-motion approach, we have%
\begin{eqnarray}
&&\left( i\omega _{n}-\epsilon _{1}\right) \left\langle \left\langle
d_{1\sigma }|d_{1\sigma }^{\dagger }\right\rangle \right\rangle  \notag \\
&=&1-t\left\langle \left\langle d_{2\sigma }|d_{1\sigma }^{\dagger
}\right\rangle \right\rangle +V_{1}^{\ast }\sum_{\mathbf{k}}\left\langle
\left\langle c_{\mathbf{k}1\sigma }|d_{1\sigma }^{\dagger }\right\rangle
\right\rangle  \notag \\
&&+V_{2}^{\ast }\sum_{\mathbf{k}}\left\langle \left\langle c_{\mathbf{k}%
2\sigma }|d_{1\sigma }^{\dagger }\right\rangle \right\rangle ,
\end{eqnarray}%
\begin{equation}
\left( i\omega _{n}-\epsilon _{\mathbf{k}1}\right) \left\langle \left\langle
c_{\mathbf{k}1\sigma }|d_{1\sigma }^{\dagger }\right\rangle \right\rangle
=V_{1}\left\langle \left\langle d_{1\sigma }|d_{1\sigma }^{\dagger
}\right\rangle \right\rangle ,
\end{equation}%
\begin{equation}
\left( i\omega _{n}-\epsilon _{\mathbf{k}2}\right) \left\langle \left\langle
c_{\mathbf{k}2\sigma }|d_{1\sigma }^{\dagger }\right\rangle \right\rangle
=V_{2}\left\langle \left\langle d_{1\sigma }|d_{1\sigma }^{\dagger
}\right\rangle \right\rangle .
\end{equation}%
Here we assume that the conduction leads have a flat and energy independent
density of states (i. e., the wide-band limit), leading to%
\begin{eqnarray*}
-\frac{V_{1}^{2}}{N}\sum_{\mathbf{k}}\frac{1}{i\omega _{n}-\epsilon _{%
\mathbf{k}1}} &\approx &i\pi \rho _{F}V_{1}^{2}\equiv i\frac{\Gamma _{1}}{2},
\\
-\frac{V_{2}^{2}}{N}\sum_{\mathbf{k}}\frac{1}{i\omega _{n}-\epsilon _{%
\mathbf{k}2}} &\approx &i\pi \rho _{F}V_{2}^{2}\equiv i\frac{\Gamma _{2}}{2},
\end{eqnarray*}%
and then%
\begin{eqnarray}
\left[ i\omega _{n}-\epsilon _{1}+\frac{i}{2}\left( \Gamma _{1}+\Gamma
_{2}\right) \right] \left\langle \left\langle d_{1\sigma }|d_{1\sigma
}^{\dagger }\right\rangle \right\rangle &&  \notag \\
+t\left\langle \left\langle d_{2\sigma }|d_{1\sigma }^{\dagger
}\right\rangle \right\rangle &=&1.
\end{eqnarray}%
By applying the similar method, we obtain the equation of motion for the
Green function $\left\langle \left\langle d_{2\sigma }|d_{1\sigma }^{\dagger
}\right\rangle \right\rangle $ as%
\begin{eqnarray}
\left[ i\omega _{n}-\epsilon _{2}+\frac{i}{2}\left( \Gamma _{3}+\Gamma
_{4}\right) \right] \left\langle \left\langle d_{2\sigma }|d_{1\sigma
}^{\dagger }\right\rangle \right\rangle &&  \notag \\
+t\left\langle \left\langle d_{1\sigma }|d_{1\sigma }^{\dagger
}\right\rangle \right\rangle &=&0.
\end{eqnarray}

\begin{figure}[tbp]
\includegraphics[width=8cm]{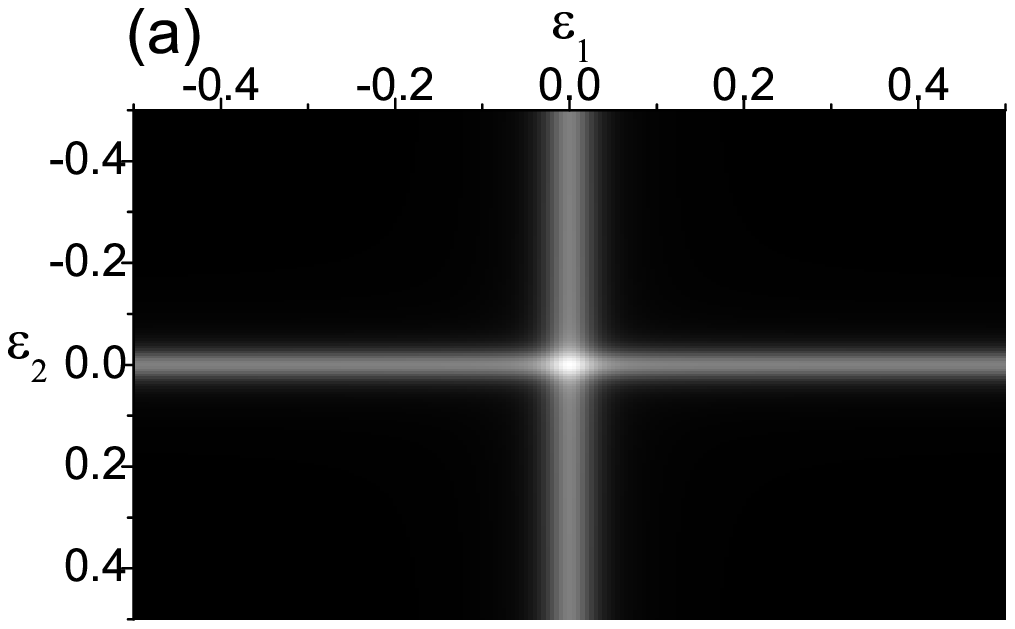} %
\includegraphics[width=8cm]{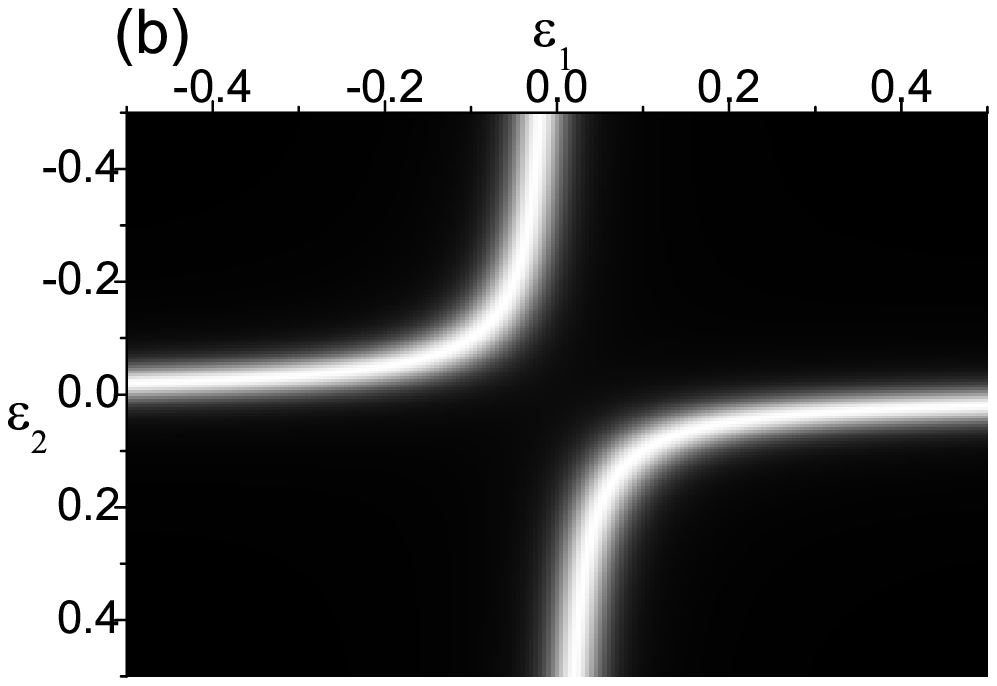}
\caption{The images of total linear conductance $G_{dot1}+G_{dot2}$ as a
function of the dot levels $\protect\epsilon_{1}$ and $\protect\epsilon_{2}$
for noninteracting ($U=0$) double dots with $\Gamma=0.01$, $k_{B}T=0.01$, $%
t=0 $ in (a), and $t=0.1$ in (b).}
\label{fig00}
\end{figure}

\begin{figure}[tbp]
\includegraphics[width=8cm]{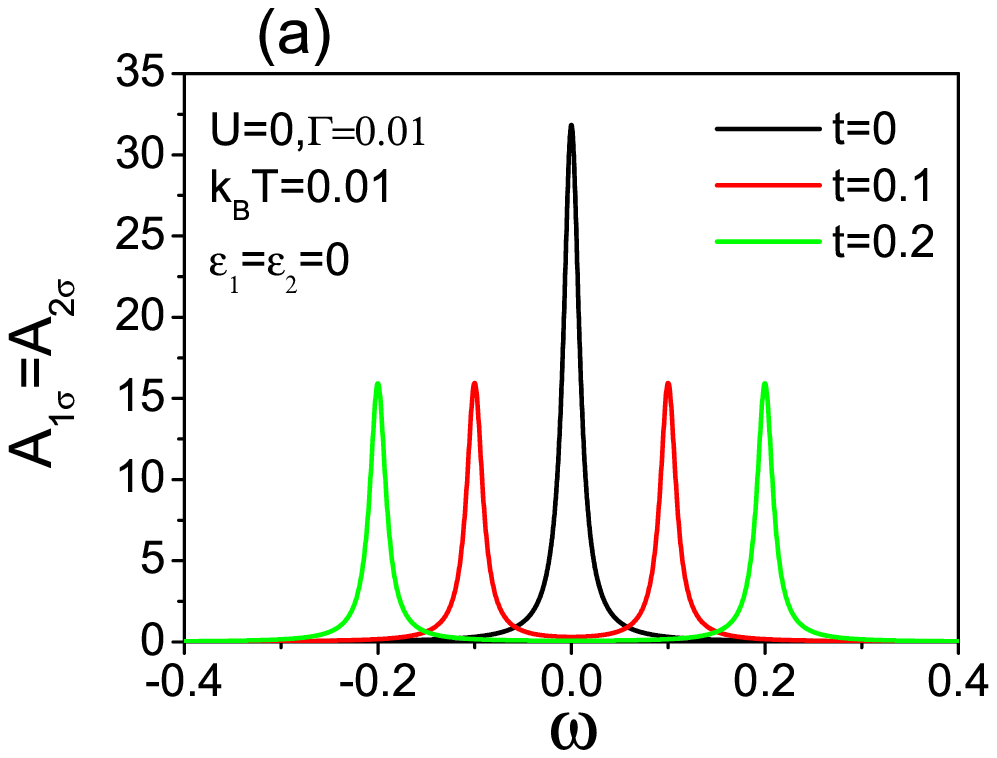} %
\includegraphics[width=8cm]{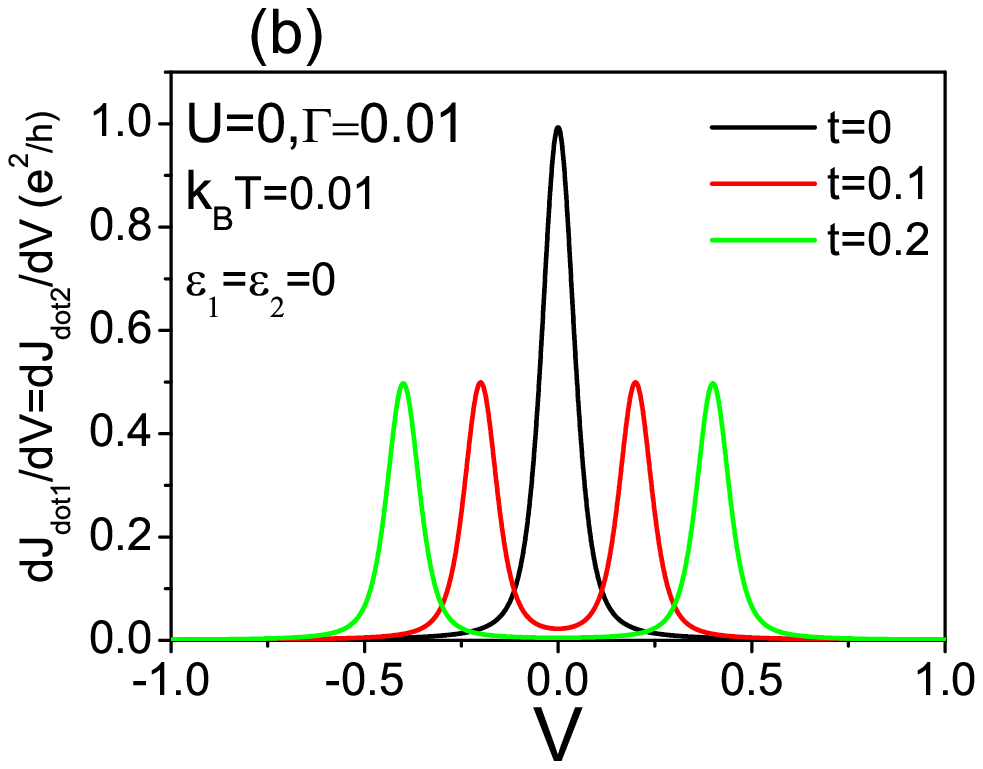}
\caption{The spectral function as a function of frequency in (a),
and the differential conductance as a function of bias voltage in
(b) with the dot-level:
$\protect\epsilon_{1}=\protect\epsilon_{2}=0$ for different
interdot tunneling couplings.} \label{fig02}
\end{figure}

From the above two equations and performing the analytic
continuations, we can deduce the retarded Green functions in the
symmetric dot-lead tunneling
coupling case,%
\begin{eqnarray*}
G_{1\sigma ,1\sigma }^{r} &=&\left( \frac{\epsilon _{+}-\epsilon _{2}}{%
\epsilon _{+}-\epsilon _{-}}\right) \frac{1}{\omega -\epsilon _{+}+i\Gamma }
\\
&&+\left( \frac{\epsilon _{2}-\epsilon _{-}}{\epsilon _{+}-\epsilon _{-}}%
\right) \frac{1}{\omega -\epsilon _{-}+i\Gamma },
\end{eqnarray*}%
\begin{eqnarray*}
G_{2\sigma ,2\sigma }^{r} &=&\left( \frac{\epsilon _{+}-\epsilon _{1}}{%
\epsilon _{+}-\epsilon _{-}}\right) \frac{1}{\omega -\epsilon _{+}+i\Gamma }
\\
&&+\left( \frac{\epsilon _{1}-\epsilon _{-}}{\epsilon _{+}-\epsilon _{-}}%
\right) \frac{1}{\omega -\epsilon _{-}+i\Gamma },
\end{eqnarray*}%
and%
\begin{eqnarray}
G_{1\sigma ,2\sigma }^{r} &=&G_{2\sigma ,1\sigma }^{r}=\left( \frac{t}{%
\epsilon _{+}-\epsilon _{-}}\right)   \notag \\
&&\times \left[ \frac{1}{\omega -\epsilon _{+}+i\Gamma
}-\frac{1}{\omega -\epsilon _{-}+i\Gamma }\right] ,
\end{eqnarray}%
where%
\begin{equation}
\epsilon _{\pm }=\frac{1}{2}\left[ \left( \epsilon _{1}+\epsilon _{2}\right)
\pm \sqrt{\left( \epsilon _{1}-\epsilon _{2}\right) ^{2}+4t^{2}}\right] ,
\label{(anti)bonding}
\end{equation}%
denote the energies of antibonding and bonding molecular resonant-tunneling
states, respectively. Note that the bonding state moves down in energy with
the interdot tunneling coupling and the antibonding state moves up. The
energy difference between the antibonding and bonding states is $\epsilon
_{+}-\epsilon _{-}=\sqrt{\left( \epsilon _{1}-\epsilon _{2}\right)
^{2}+4t^{2}}$. When the levels of two dots cross, i. e. $\epsilon
_{1}=\epsilon _{2}$, one has an anticrossing of $\epsilon _{+}$ and $%
\epsilon _{-}$ with the minimum antibonding-bonding energy
difference $2t$. For large dot-level energy difference, the
eigenenergies of tunneling-coupled double dots
approach to the energy levels of uncoupled dots, $%
\epsilon _{1}$ and $\epsilon _{2}$.

From the retarded Green functions, the spectral function of dot 1 is found to be%
\begin{eqnarray}
\mathcal{A}_{1\sigma }\left( \omega \right) &=&\frac{1}{\pi }\left[ \left(
\frac{\epsilon _{+}-\epsilon _{2}}{\epsilon _{+}-\epsilon _{-}}\right) \frac{%
\Gamma }{\left( \omega -\epsilon _{+}\right) ^{2}+\Gamma ^{2}}\right.  \notag
\\
&&\left. +\left( \frac{\epsilon _{2}-\epsilon _{-}}{\epsilon _{+}-\epsilon
_{-}}\right) \frac{\Gamma }{\left( \omega -\epsilon _{-}\right) ^{2}+\Gamma
^{2}}\right] ,  \label{SF1}
\end{eqnarray}%
and changing the index $1\leftrightarrow 2$ gives the spectral function of
dot 2. The spectral functions of dot 1 and 2 display two Lorentzian
resonances at the same positions: $\epsilon _{\pm }$, while the ratio of
their heights is $\left( \epsilon _{+}-\epsilon _{2}\right) /\left( \epsilon
_{2}-\epsilon _{-}\right) $ for the dot 1, and $\left( \epsilon
_{+}-\epsilon _{1}\right) /\left( \epsilon _{1}-\epsilon _{-}\right) $ for
the dot 2.

In Fig. \ref{fig00} we show the images of the total linear conductance $%
G_{dot1}+G_{dot2}$, obtained by Eq. (\ref{linear-conductance}),
versus changes of levels $\epsilon _{1}$ and $\epsilon _{2}$ of
two dots. The bright regions correspond to high conductance and
the dark regions to low
conductance. For completely decoupled dots ($t=0$) as in Fig. \ref{fig00}%
(a), tuning the level of one dot changes the charge on this dot
without affecting the charge on the other dot. For nonzero
interdot tunneling coupling, a similar ''anticrossing'' occurs at
$\epsilon _{1}=\epsilon _{2}$ in the diagram of conductance [see
Fig. \ref{fig00}(b))].

In Fig. \ref{fig02}, the spectral function and the differential conductance
of dots are displayed for $\epsilon _{1}=\epsilon _{2}=0$ and different
interdot tunneling couplings. In the absence of interdot tunneling, the
electron transports through each dot independently, which exhibits a
resonant peak when the dot levels match the chemical potential of the leads.
Increasing the interdot tunneling leads to the splitting of resonant peak at
the positions $\epsilon _{\pm }$, corresponding to the bonding and
antibonding molecular states of the DQD system. Tuning the bias voltage
gives rise to the resonances in the differential conductance when the
resonant peaks in the spectral function enter the region between the
chemical potentials of leads.
\begin{figure}[tbp]
\includegraphics[width=8cm]{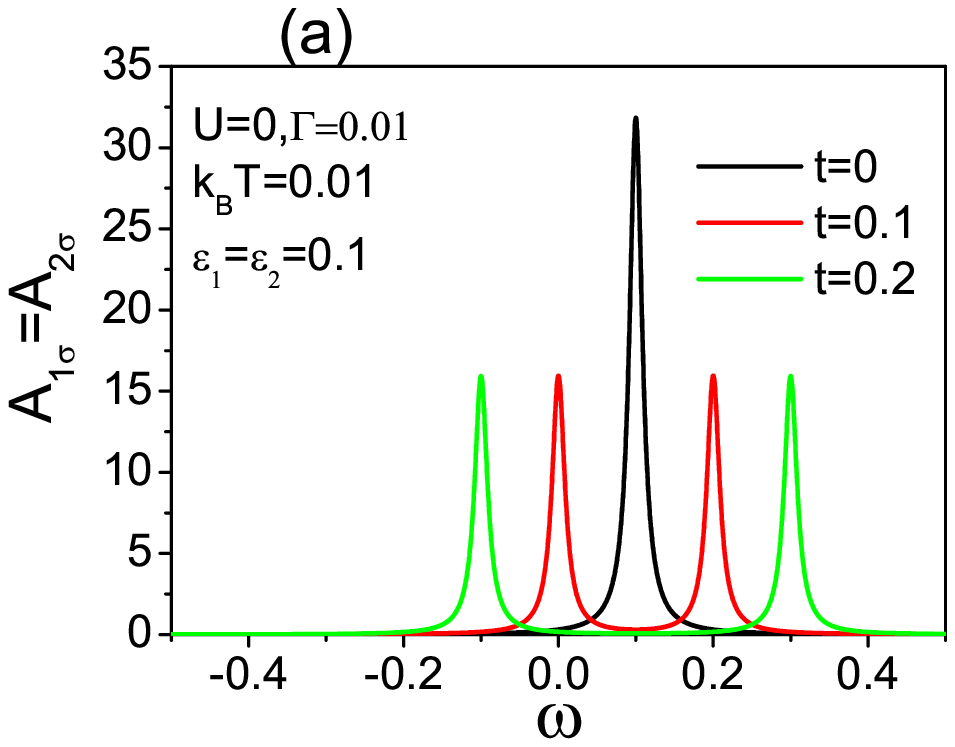} %
\includegraphics[width=8cm]{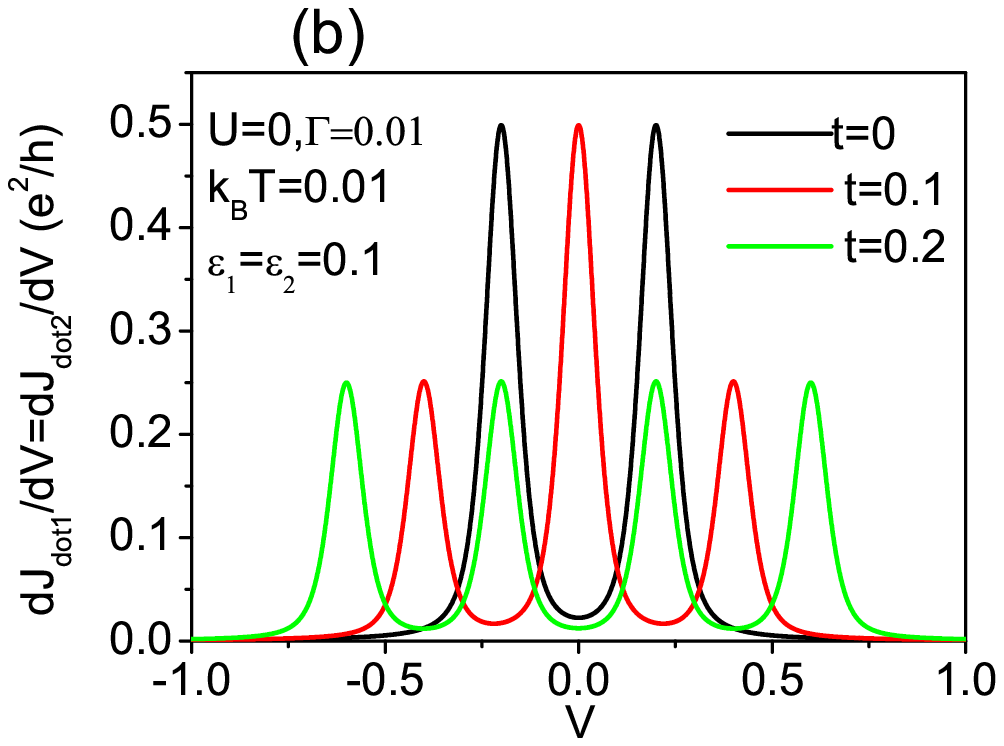}
\caption{The spectral function (a) and the differential conductance (b) with
the dot-level: $\protect\epsilon _{1}=\protect\epsilon _{2}=0.1$ for
different interdot tunneling coupling.}
\label{fig03}
\end{figure}

We also show the spectral function and the differential
conductance of dots along the direction $\epsilon _{1}=\epsilon
_{2}$ for different interdot tunneling couplings in Fig.
\ref{fig03}. Compared to Fig. \ref{fig02}, the spectral function
is shifted to $\epsilon _{1}=\epsilon _{2}=0.1$. The asymmetric
spectral function along $\omega $ leads to the four peaks
structure in the differential conductance as a function of the
bias voltage for the general configuration of dot level and the
finite interdot tunneling coupling [see the green line of Fig.
\ref{fig03}(b)]. One interesting observation is that under the
condition $t=\epsilon _{1}=\epsilon _{2}$, this general four peaks
structure reduces to the three peaks structure [see the red line of Fig. \ref%
{fig03}(b)], while in this case the antibonding energy shifts to
zero [see the red line of corresponding spectral function in Fig.
\ref{fig03}(a)]. This feature offers a possible method to measure
the tunneling coupling between dots in experiments.
\begin{figure}[tbp]
\includegraphics[width=9cm]{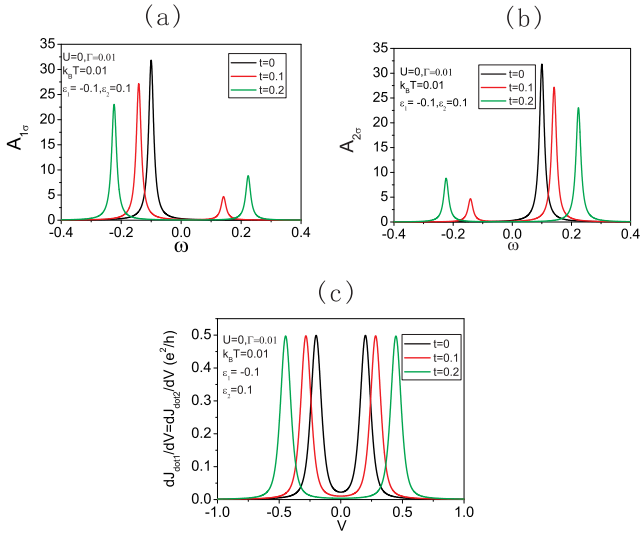}
\caption{The spectral function of the dot 1 in (a) and the dot 2 in (b) and
the differential conductance in (c) with the dot-level: $\protect\epsilon %
_{1}=-0.1$,$\protect\epsilon _{2}=0.1$ for different interdot tunneling
coupling.}
\label{fig04}
\end{figure}
Fig. \ref{fig04} displays the spectral function and the differential
conductance along the direction $\epsilon _{1}=-\epsilon _{2}=-\epsilon _{d}$
for different interdot tunneling couplings. In this case, from Eq. (\ref%
{(anti)bonding}) one can easily see that the bonding and antibonding
energies reduce to $\epsilon _{\pm }=\pm \left( \epsilon _{d}+t\right) $,
which yields two resonant peaks in the spectral function for finite interdot
tunneling coupling, while the heights of these two peaks are different for
the ratio of heights is $t/\left( 2\epsilon _{d}+t\right) $ for dot 1 and $%
\left( 2\epsilon _{d}+t\right) /t$ for dot 2. As these two resonant peaks
sit at the positions with the same absolute value, there are only two
resonances in the differential conductance compared to the four resonances
in the case $\epsilon _{1}=\epsilon _{2}$ (see Fig. \ref{fig04}).

\begin{figure}[tbp]
\includegraphics[width=8cm]{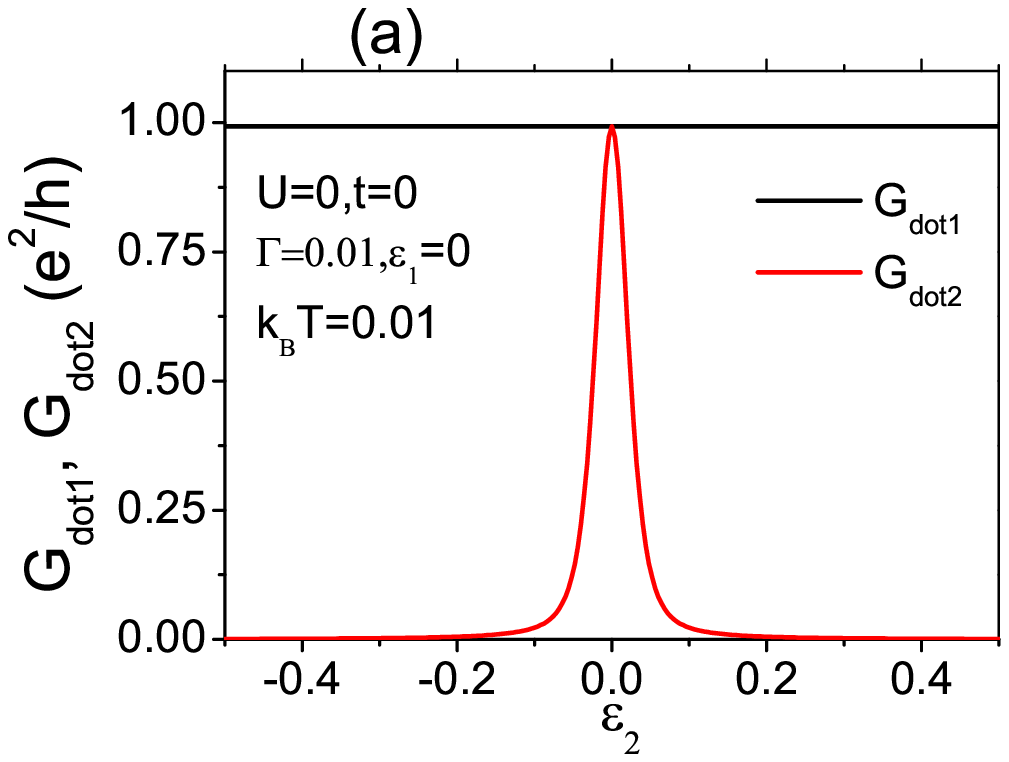} %
\includegraphics[width=8cm]{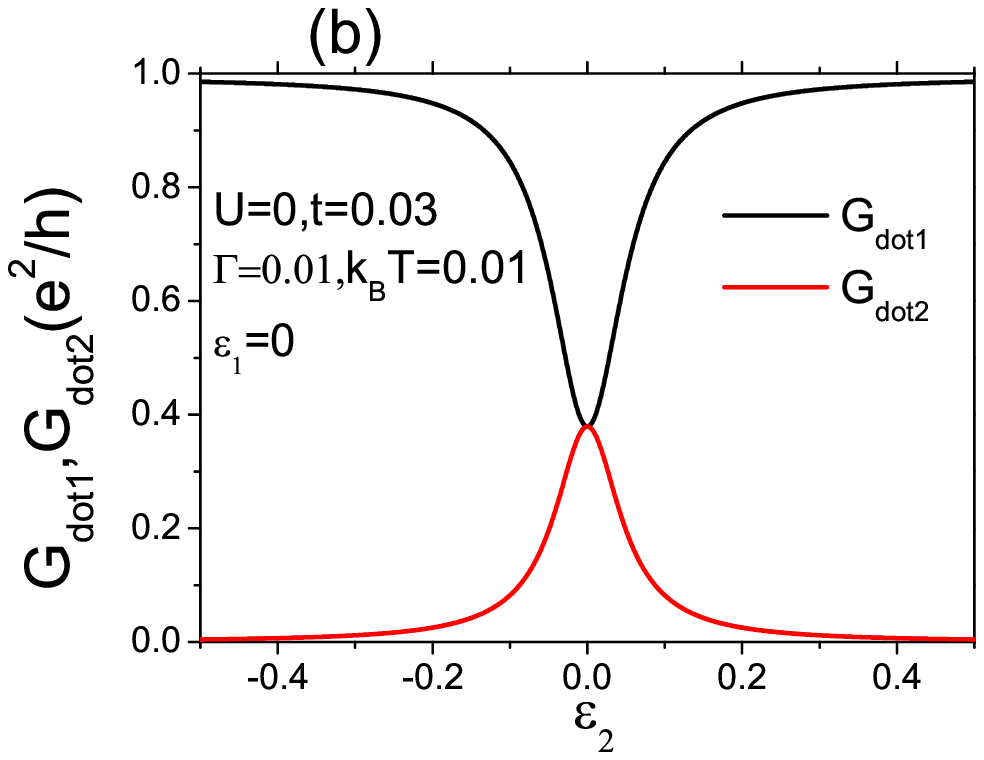} %
\caption{The linear conductance of dot 1 and 2 as a function of dot level of
one dot $\protect\epsilon_{2}$ by setting another dot's level as $\protect%
\epsilon_{1}=0$ without (a) and with (b) the interdot tunneling coupling.}
\label{fig0a}
\end{figure}

\begin{figure}[tbp]
\includegraphics[width=9cm]{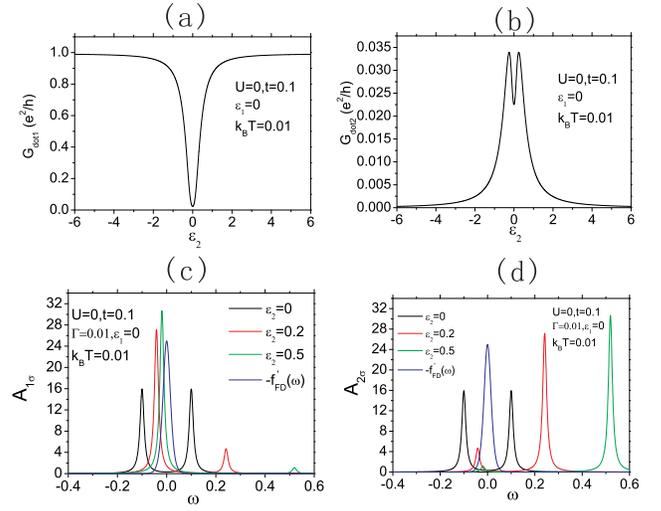} %
\caption{The linear conductance of dot 1 (a) and 2 (b) as a function of dot
level of one dot $\protect\epsilon _{2}$ by setting another dot's level as $%
\protect\epsilon _{1}=0$ when increasing the interdot tunneling coupling,
and the associated spectral functions for different $\protect\epsilon _{2}$
[(c) and (d)].}
\label{fig0b}
\end{figure}

In Fig. \ref{fig0a}, the linear conductance of each dot as a
function of one dot's level is displayed by setting another dot's
level. In the absence of the interdot tunneling coupling, electron
transports independently through each dot, which shows one
constant linear conductance of dot 1 and one resonant peak in the
linear conductance of dot 2 when $\epsilon_{2}$ matches the
chemical potential of leads. In the presence of interdot tunneling
coupling, the linear conductance of dot 1 shows a dip at the same
position of $\epsilon _{2}$, where the linear conductance of dot 2
exhibits a resonant peak.

An interesting feature is shown in Fig. \ref{fig0b} when increasing the
interdot tunneling coupling: the resonant peak in the linear conductance of
dot 2 is suppressed into two splitting peaks. Both the dip in $G_{dot1}$ and
the splitting peaks in $G_{dot2}$ could be understood with the help of the
associated spectral function of each dot for different dot level $\epsilon
_{2}$. The blue line in $A_{1(2)\sigma }$ of Fig. \ref{fig0b} shows $%
-\partial f_{FD}/\partial \epsilon $ in the formula of linear
conductance Eq. (\ref{linear-conductance}), which sets the window
of temperature. It is noted that the spectral functions of both
dots will include two peaks, corresponding to the bonding and
antibonding states at finite interdot tunneling coupling, while
the heights of these two peaks are different. For the spectral
function of dot 1 ($A_{1\sigma }$), when increasing the dot level
$\epsilon _{2}$, the peak corresponding to bonding state $\epsilon
_{-} $, which is smaller than the minimum of $\epsilon _{1}$ and
$\epsilon _{2}$, moves to the window of temperature with the
increasing of its height. This will enhance the linear conductance
of dot 1 as increasing $\epsilon _{2}$ from $\epsilon _{2}=0$, and
the linear conductance saturates to a constant value as the
bonding-state peak completely enter into the window of temperature
[see Fig. \ref{fig0b}(a)]. For the spectral function of dot 2
($A_{2\sigma }$), as increasing the dot level $\epsilon _{2}$, the
bonding-state peak moves into the window of temperature, but the
height of this peak decreases. This leads to a maximum in linear
conductance of dot 2 [see Fig. \ref{fig0b}(b)]. While this maximum
in the linear conductance of dot 2 could not be observed in the
small interdot coupling case (as shown in Fig. \ref{fig0a}) or in
the low temperature case, because the bonding-state peak is too
close to $\omega =0$ or the window of temperature is too narrow.

\section{The interacting dots}

In this section we study the transport phenomena through
tunneling-coupled DQD system with intradot Coulomb interaction
with the help of equation-of-motion approach. This method consists
of differentiating the Green function with respect to time,
thereby generating higher-order Green functions in the presence of
intradot on-site Coulomb interaction which eventually have to be
closed. Here we apply the Hartree-Fock approximation to truncate
the higher-order Green functions, which includes the contributions
of higher order electron tunneling processes. It is known that the
Hartree-Fock approximation captures the correct qualitative
feature of physics of single quantum dot in the Coulomb blockade
regime, valid in the higher temperature regime.

By applying the equation-of-motion approach and the Hartree-Fock
approximation to truncate the higher-order Green functions (for
detailed calculation see the Appendix A), we obtain the Green
functions for dot electrons as
\begin{widetext}
\begin{equation}
\left\langle \left\langle d_{1\sigma }|d_{1\sigma }^{\dagger
}\right\rangle \right\rangle =\frac{F_{1} \left[ i\omega
_{n}-\epsilon _{2}+F_{2} \frac{i}{2}\left( \Gamma _{3}+\Gamma
_{4}\right) \right] }{\left[ i\omega _{n}-\epsilon _{1}+F_{1}
\frac{i\left( \Gamma _{1}+\Gamma _{2}\right) }{2}\right] %
\left[ i\omega _{n}-\epsilon _{2}+F_{2} \frac{i\left( \Gamma
_{3}+\Gamma _{4}\right) }{2}\right] -t^{2}F_{1} F_{2} },
\label{GF11}
\end{equation}
\begin{equation}
\left\langle \left\langle d_{2\sigma }|d_{2\sigma }^{\dagger
}\right\rangle \right\rangle =\frac{F_{2} \left[ i\omega
_{n}-\epsilon _{1}+F_{1} \frac{i}{2}\left( \Gamma _{1}+\Gamma
_{2}\right) \right] }{\left[ i\omega _{n}-\epsilon _{1}+F_{1}
\frac{i\left( \Gamma _{1}+\Gamma _{2}\right) }{2}\right] %
\left[ i\omega _{n}-\epsilon _{2}+F_{2} \frac{i\left( \Gamma
_{3}+\Gamma _{4}\right) }{2}\right] -t^{2}F_{1} F_{2} },
\label{GF22}
\end{equation}
and
\begin{equation}
\left\langle \left\langle d_{2\sigma }|d_{1\sigma }^{\dagger
}\right\rangle \right\rangle =\left\langle \left\langle d_{1\sigma
}|d_{2\sigma }^{\dagger
}\right\rangle \right\rangle\frac{-tF_{1} F_{2} }{%
\left[ i\omega _{n}-\epsilon _{1}+F_{1} \frac{i\left( \Gamma
_{1}+\Gamma _{2}\right) }{2}\right] \left[ i\omega _{n}-\epsilon
_{2}+F_{2} \frac{i\left( \Gamma _{3}+\Gamma _{4}\right)
}{2}\right] -t^{2}F_{1} F_{2} }, \label{GF12}
\end{equation}
\end{widetext}
where
\begin{eqnarray}
F_{1}\left( i\omega _{n},\epsilon _{1},\left\langle n_{1\overline{\sigma }%
}\right\rangle \right) &=&1+\frac{U\left\langle n_{1\overline{\sigma }%
}\right\rangle }{i\omega _{n}-\epsilon _{1}-U},  \notag \\
F_{2}\left( i\omega _{n},\epsilon _{2},\left\langle n_{2\overline{\sigma }%
}\right\rangle \right) &=&1+\frac{U\left\langle n_{2\overline{\sigma }%
}\right\rangle }{i\omega _{n}-\epsilon _{2}-U}.
\end{eqnarray}
Performing the analytic continuations, we can deduce the retarded
Green functions for dot electrons in the tunneling-coupled DQD
system with intradot on-site Coulomb interaction:
$G_{i\sigma,j\sigma}^{r}$, where $i, j= 1, 2$ corresponding dot 1
and 2 respectively. The occupation numbers are subjected to the
self-consistency condition:
\begin{equation}
\left\langle n_{i\sigma }\right\rangle =-i\int \frac{d\omega }{2\pi }%
G_{i\sigma ,i\sigma }^{<}\left( \omega \right) ,
\label{self-consistency}
\end{equation}
where $i=1$ and $2$.

Next we calculate the Keldysh less Green functions $G_{i\sigma
,i\sigma }^{<}\left( \omega \right) $, which is needed in the
self-consistent equation Eq. (\ref{self-consistency}) for the dot
occupation numbers. Under the same Hartree-Fock approximation as
in the retarded Green functions, the equation of motion approach
gives that the Keldysh less Green functions satisfy (for detailed
calculation see the Appendix B)
\begin{eqnarray}
G_{1\sigma ,1\sigma }^{<} &=&i\left[ \Gamma _{1}f_{1}\left( \omega
\right) +\Gamma _{2}f_{2}\left( \omega \right) \right] G_{1\sigma
,1\sigma
}^{r}G_{1\sigma ,1\sigma }^{a}  \notag \\
&&+i\left[ \Gamma _{3}f_{3}\left( \omega \right) +\Gamma
_{4}f_{4}\left( \omega \right) \right] G_{2\sigma ,1\sigma
}^{r}G_{2\sigma ,1\sigma }^{a}, \label{G_less_11}
\end{eqnarray}%
for dot 1, and%
\begin{eqnarray}
G_{2\sigma ,2\sigma }^{<} &=&i\left[ \Gamma _{1}f_{1}\left( \omega
\right) +\Gamma _{2}f_{2}\left( \omega \right) \right] G_{1\sigma
,2\sigma
}^{r}G_{1\sigma ,2\sigma }^{a}  \notag \\
&&+i\left[ \Gamma _{3}f_{3}\left( \omega \right) +\Gamma
_{4}f_{4}\left( \omega \right) \right] G_{2\sigma ,2\sigma
}^{r}G_{2\sigma ,2\sigma }^{a}, \label{G_less_22}
\end{eqnarray}
for dot 2.

\begin{figure}[tbp]
\includegraphics[width=7cm]{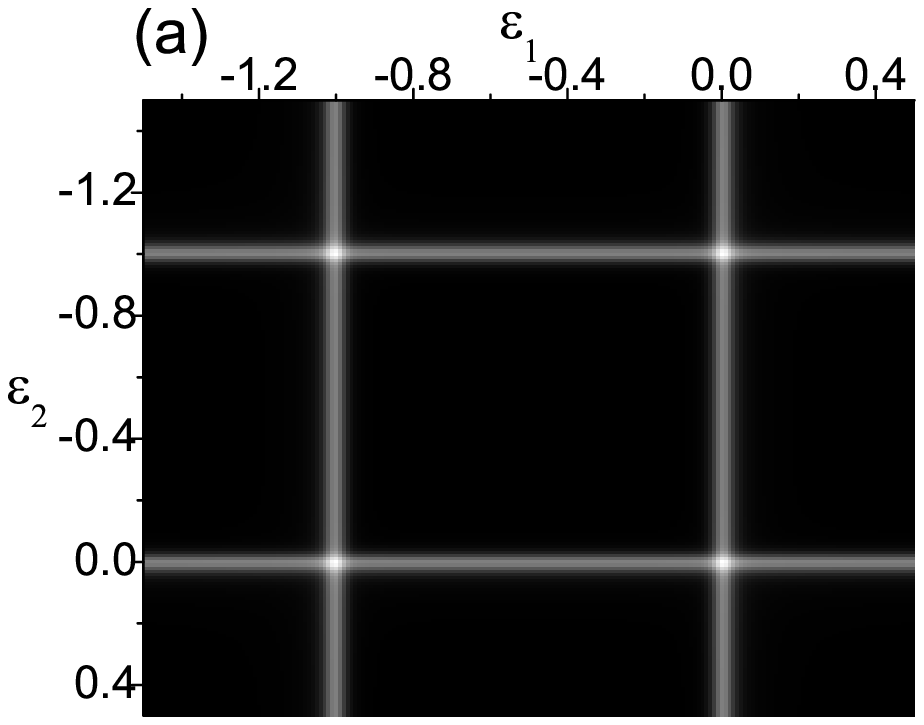} %
\includegraphics[width=7cm]{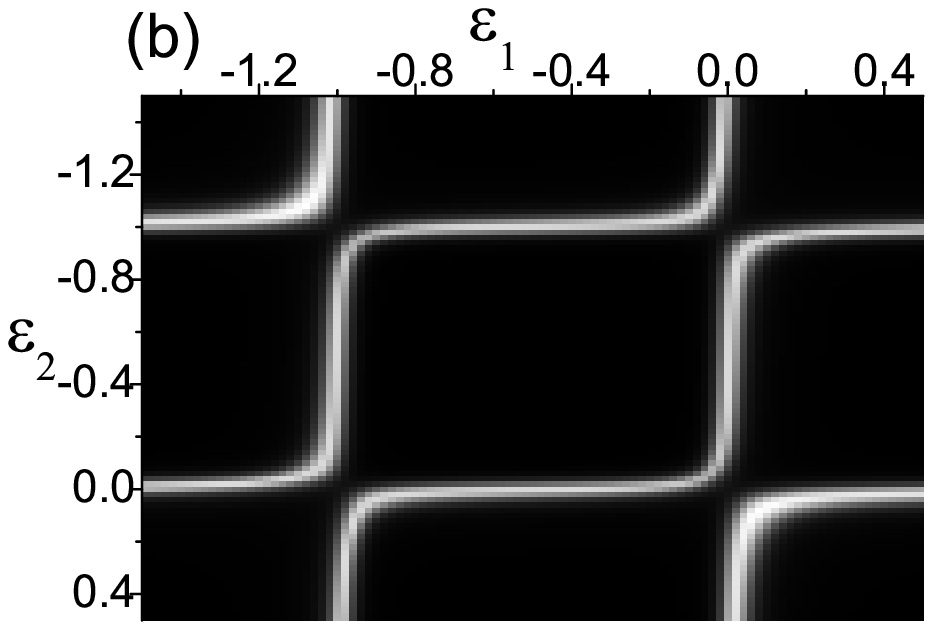}
\caption{The images of total linear conductance
$G_{dot1}+G_{dot2}$ as a
function of the dot levels $\protect\epsilon _{1}$ and $\protect\epsilon %
_{2} $ for interacting ($U=1$) double dots with $\Gamma =0.01$,
$k_{B}T=0.01$ $t=0 $ in (a), and $t=0.1$ in (b).} \label{fig05}
\end{figure}

\begin{figure}[tbp]
\includegraphics[width=8cm]{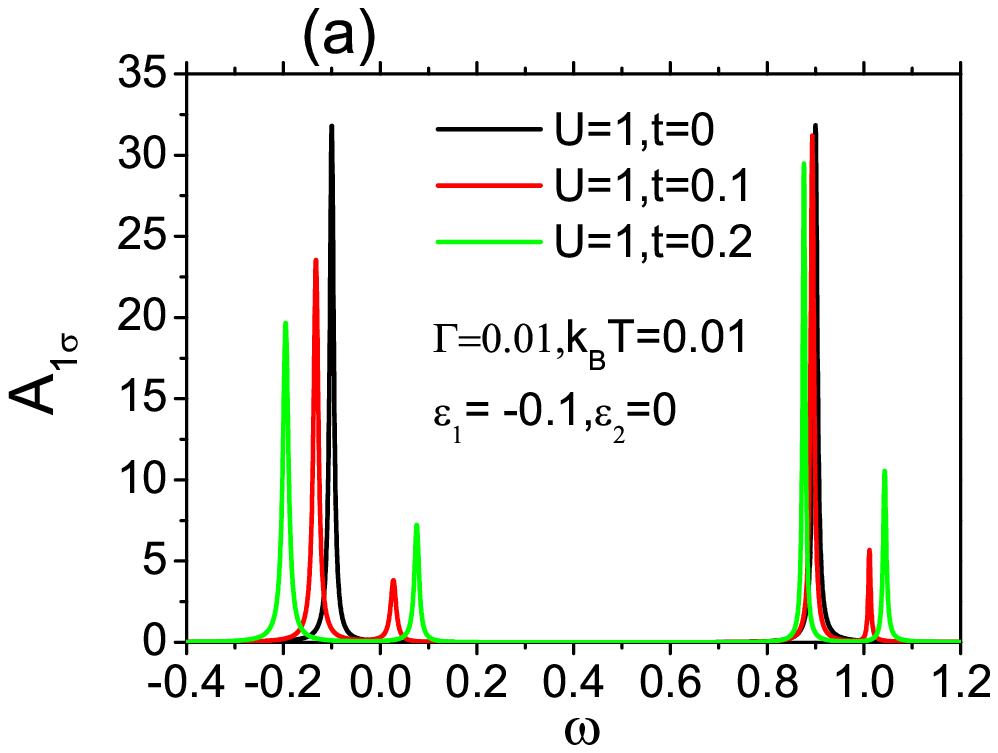} %
\includegraphics[width=8cm]{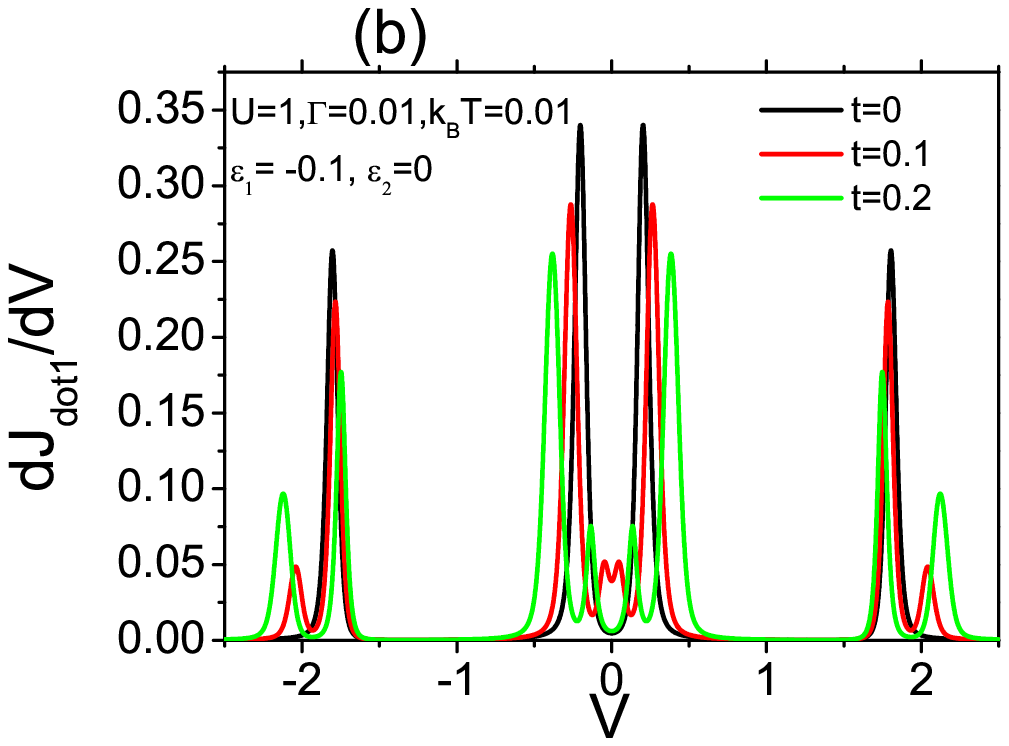}
\caption{The spectral function (a) and the differential
conductance (b) of the dot 1 in the DQD system with the intradot
Coulomb interaction in the
general case of the dot-level: $\protect\epsilon _{1}=-0.1$, $\protect%
\epsilon _{2}=0$ for different interdot tunneling coupling.}
\label{fig07}
\end{figure}
\begin{figure}[tbp]
\includegraphics[width=8cm]{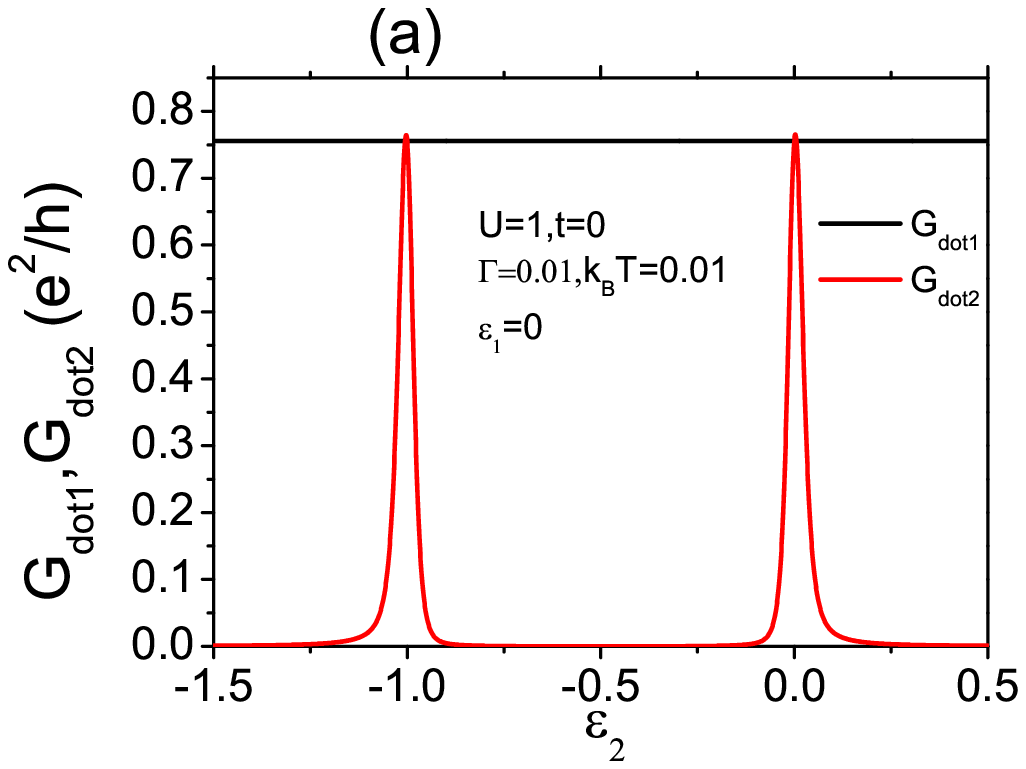} %
\includegraphics[width=8cm]{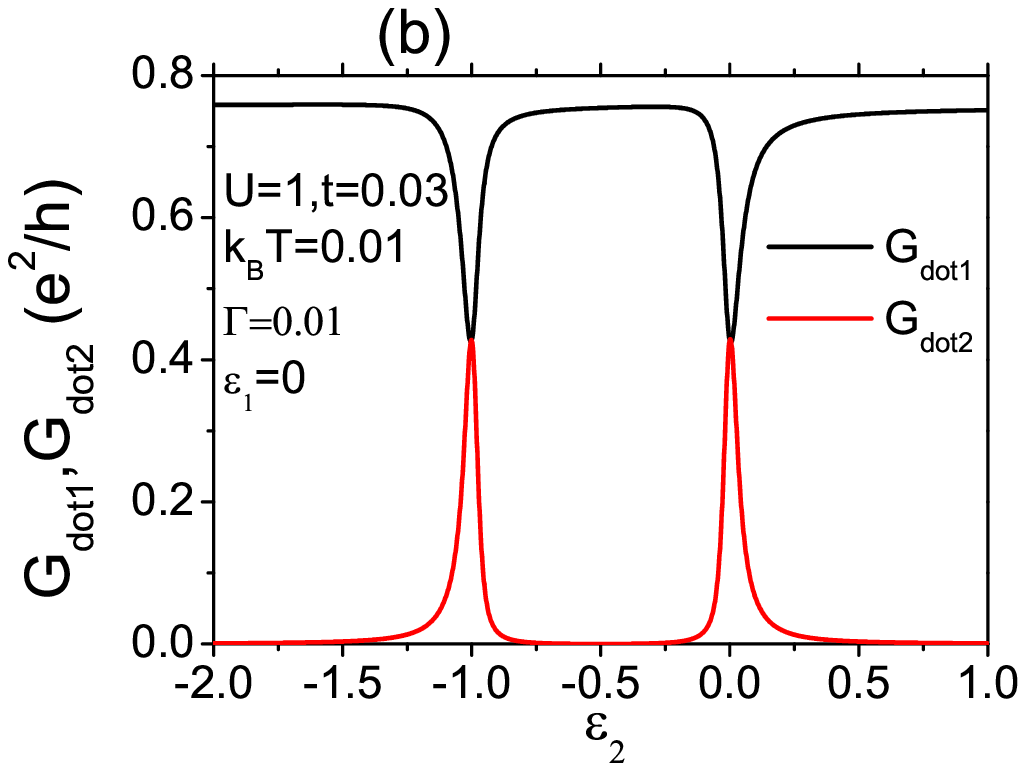}
\caption{The linear conductance of dot 1 and 2 with the
intradot-dot Coulomb interaction as a function of dot level of one
dot $\protect\epsilon _{2}$ by setting another dot's level as
$\protect\epsilon _{1}=0$ without [in (a)] and with [in (b)] the
interdot tunneling coupling.} \label{fig0c}
\end{figure}

Fig. \ref{fig05} shows the images of the total linear conductance $%
G_{dot1}+G_{dot2}$ of DQD with the intradot on-site Coulomb interaction
versus the changes of the dot-level $\epsilon_{1}$ and $%
\epsilon_{2}$. Without interdot tunneling coupling, the images of
conductance shows a square lattice structure [see Fig.
\ref{fig05}(a)]. In the presence of both the interdot tunneling
coupling and the intradot Coulomb
interaction, the "anticrossing" occurs at the positions of $\epsilon_{1}$, $%
\epsilon_{2}$, $\epsilon_{1}+U$, and $\epsilon_{2}+U$ [see Fig. \ref{fig05}%
(b)].

The effects of the intradot Coulomb interaction and the interdot
tunneling coupling can be clearly shown in the spectral functions
of quantum dots. Fig. \ref{fig07} is the spectral function and the
differential conductance of the dot 1. In the absence of the
interdot tunneling coupling, the intradot Coulomb interaction
leads to two peaks at $\epsilon _{1(2)}$ and $\epsilon _{1(2)}+U$
in the spectral function, in agreement with the well-known result
of Coulomb blockade at higher temperatures. These two peaks split
in the presence of the interdot tunneling coupling, and the
spectral function displays a four peak structure. The differential
conductance exhibits the resonance peaks when the spectral
function peaks enter the window of bias voltage. Due to the small
size of QD, the Coulomb charging energy $U$ becomes the largest
energy scale. In the case $U>t$, the peaks in both the spectral
function and the linear conductance form two distinct groups
separated by roughly the intradot Coulomb interaction $U$, while
in each group there are two peaks separated by roughly the
interdot tunneling coupling $t$.

In Fig. \ref{fig0c} we show the linear conductance of each dot with the
intradot Coulomb interaction as a function of one dot-level by setting
another dot-level. In the absence of interdot tunneling coupling, by setting
$\epsilon _{1}$ and tuning $\epsilon _{2}$, $G_{dot1}$ remains constant and $%
G_{dot2}$ shows two resonant peaks when the levels $\epsilon _{2}$ and $%
\epsilon _{2}+U$ match the chemical potential of leads. Due to the
interdot tunneling coupling, the electron transport through one
dot is strongly influenced by the electron transport through
another dot: $G_{dot1}$ displays two dips at the corresponding
resonant positions of $G_{dot2}$. This feature provides a possible
experimental method to observe the correlated electron transport
phenomena in the real interacting DQD system with strong intradot
tunneling coupling at high temperatures.

\section{Summary}

In this work we have investigated theoretically the electron
transport through parallel tunneling-coupled DQD system with the
special configuration that two leads are connected to each dot. By
applying the Keldysh nonequilibrium Green function technique and
the equation-of-motion approach, we have obtained the spectral and
conductance spectra of DQD with and without the intradot on-site
Coulomb interaction, respectively.

The main results are summarized as following:

1) In the absence of intradot Coulomb interaction, the exact results of the
spectral function and the conductance and obtained, showing resonant peaks
at the positions of antibonding and bonding molecular states for finite
interdot tunneling couplings, while the heights of these resonant peaks
depend on the levels of two dots.

2) Tuning the dot levels results in different structure of the differential
conductance, e. g., under the condition $\epsilon_{1}=\epsilon_{2}=t$, the
general four-peak structure in the differential conductance can reduce to
the three-peak structure.

3) For DQD with the intradot Coulomb interaction, the Hartree-Fock
approximation has been applied to calculate both the retarded and the
Keldysh less Green functions, which gives the transport properties at high
temperatures.

4) As the interdot Coulomb interaction $U$ becomes the largest energy scale
of quantum dot in the single-electron tunneling regime, the peaks in the
spectral function and the conductance form two groups separated by roughly $%
U $, and the interdot tunneling coupling $t$ induces the splitting of peaks
in each group.

5) The linear conductance of each dot shows the phenomena of
correlated electron transport through tunneling-coupled DQD
system, which could be observed with the present experimental
techniques.

\begin{acknowledgments}

R. L\"{u} is supported by the MOE of China (Grant No.200221). G.M. Zhang is
supported by NSF-China (Grant No.10125418) and the Special Fund for Major
State Basic Research Projects of China (Grant No.G2000067107).

\end{acknowledgments}

\appendix

\section{The Evaluation of Retarded Green Functions}

In this Appendix, we give the detailed evaluation of the retarded
Green functions for dot electrons in the tunneling-coupled DQD
with intradot on-site Coulomb interaction based on the
equation-of-motion approach. Its equation of motion is yielded as
\begin{eqnarray}
&&\left( i\omega _{n}-\epsilon _{1}\right) \left\langle
\left\langle
d_{1\sigma }|d_{1\sigma }^{\dagger }\right\rangle \right\rangle  \notag \\
&=&1-t\left\langle \left\langle d_{2\sigma }|d_{1\sigma }^{\dagger
}\right\rangle \right\rangle +U\left\langle \left\langle n_{1\overline{%
\sigma }}d_{1\sigma }|d_{1\sigma }^{\dagger }\right\rangle
\right\rangle
\notag \\
&&+V_{1}^{\ast }\sum_{\mathbf{k}}\left\langle \left\langle c_{\mathbf{k}%
1\sigma }|d_{1\sigma }^{\dagger }\right\rangle \right\rangle
+V_{2}^{\ast }\sum_{\mathbf{k}}\left\langle \left\langle
c_{\mathbf{k}2\sigma }|d_{1\sigma }^{\dagger }\right\rangle
\right\rangle ,
\end{eqnarray}%
\begin{equation}
\left( i\omega _{n}-\epsilon _{\mathbf{k}1}\right) \left\langle
\left\langle c_{\mathbf{k}1\sigma }|d_{1\sigma }^{\dagger
}\right\rangle \right\rangle =V_{1}\left\langle \left\langle
d_{1\sigma }|d_{1\sigma }^{\dagger }\right\rangle \right\rangle ,
\end{equation}%
\begin{equation}
\left( i\omega _{n}-\epsilon _{\mathbf{k}2}\right) \left\langle
\left\langle c_{\mathbf{k}2\sigma }|d_{1\sigma }^{\dagger
}\right\rangle \right\rangle =V_{2}\left\langle \left\langle
d_{1\sigma }|d_{1\sigma }^{\dagger }\right\rangle \right\rangle ,
\end{equation}%
where the higher-order Green function $\left\langle \left\langle n_{1%
\overline{\sigma }}d_{1\sigma }|d_{1\sigma }^{\dagger
}\right\rangle \right\rangle $ is generated. The equation of
motion of $\left\langle \left\langle n_{1\overline{\sigma
}}d_{1\sigma }|d_{1\sigma }^{\dagger }\right\rangle \right\rangle
$ can be further deduced to
\begin{eqnarray}
&&\left( i\omega _{n}-\epsilon _{1}-U\right) \left\langle \left\langle n_{1%
\overline{\sigma }}d_{1\sigma }|d_{1\sigma }^{\dagger
}\right\rangle
\right\rangle  \notag \\
&&=\left\langle n_{1\overline{\sigma }}\right\rangle -t\left(
\left\langle \left\langle n_{1\overline{\sigma }}d_{2\sigma
}|d_{1\sigma }^{\dagger }\right\rangle \right\rangle +\left\langle
\left\langle d_{1\overline{\sigma }}^{\dagger
}d_{2\overline{\sigma }}d_{1\sigma }|d_{1\sigma }^{\dagger
}\right\rangle \right\rangle \right.  \notag \\
&&\left. -\left\langle \left\langle d_{2\overline{\sigma }}^{\dagger }d_{1%
\overline{\sigma }}d_{1\sigma }|d_{1\sigma }^{\dagger
}\right\rangle
\right\rangle \right)  \notag \\
&&-\sum_{\mathbf{k}}\left( V_{1}\left\langle \left\langle c_{\mathbf{k}1%
\overline{\sigma }}^{\dag }d_{1\overline{\sigma }}d_{1\sigma
}|d_{1\sigma }^{\dagger }\right\rangle \right\rangle -V_{1}^{\ast
}\left\langle \left\langle n_{1\overline{\sigma
}}c_{\mathbf{k}1\sigma }|d_{1\sigma
}^{\dagger }\right\rangle \right\rangle \right.  \notag \\
&&\left. -V_{1}^{\ast }\left\langle \left\langle d_{1\overline{\sigma }%
}^{\dag }c_{\mathbf{k}1\overline{\sigma }}d_{1\sigma }|d_{1\sigma
}^{\dagger }\right\rangle \right\rangle \right) .
\end{eqnarray}%
Making the Hartree-Fock approximation to decouple the higher-order
Green functions in the above equation:
\begin{eqnarray}
\left\langle \left\langle d_{1\overline{\sigma }}^{\dagger }d_{2\overline{%
\sigma }}d_{1\sigma }|d_{1\sigma }^{\dagger }\right\rangle
\right\rangle &\simeq &\left\langle d_{1\overline{\sigma
}}^{\dagger }d_{2\overline{\sigma }}\right\rangle \left\langle
\left\langle d_{1\sigma }|d_{1\sigma }^{\dagger
}\right\rangle \right\rangle ,  \notag \\
\left\langle \left\langle d_{2\overline{\sigma }}^{\dagger }d_{1\overline{%
\sigma }}d_{1\sigma }|d_{1\sigma }^{\dagger }\right\rangle
\right\rangle &\simeq &\left\langle d_{2\overline{\sigma
}}^{\dagger }d_{1\overline{\sigma }}\right\rangle \left\langle
\left\langle d_{1\sigma }|d_{1\sigma }^{\dagger
}\right\rangle \right\rangle ,  \notag \\
\left\langle \left\langle c_{\mathbf{k}1\overline{\sigma }}^{\dag }d_{1%
\overline{\sigma }}d_{1\sigma }|d_{1\sigma }^{\dagger
}\right\rangle \right\rangle &\simeq &\left\langle
c_{\mathbf{k}1\overline{\sigma }}^{\dag }d_{1\overline{\sigma
}}\right\rangle \left\langle \left\langle d_{1\sigma
}|d_{1\sigma }^{\dagger }\right\rangle \right\rangle ,  \notag \\
\left\langle \left\langle d_{1\overline{\sigma }}^{\dag }c_{\mathbf{k}1%
\overline{\sigma }}d_{1\sigma }|d_{1\sigma }^{\dagger
}\right\rangle
\right\rangle &\simeq &\left\langle d_{1\overline{\sigma }}^{\dag }c_{%
\mathbf{k}1\overline{\sigma }}\right\rangle \left\langle
\left\langle d_{1\sigma }|d_{1\sigma }^{\dagger }\right\rangle
\right\rangle .
\end{eqnarray}
and considering the facts: $V_{1}=V_{1}^{\ast }$, $\left\langle d_{1\overline{%
\sigma }}^{\dagger }d_{2\overline{\sigma }}\right\rangle =\left\langle d_{2%
\overline{\sigma }}^{\dagger }d_{1\overline{\sigma }}\right\rangle $, and $%
\left\langle c_{\mathbf{k}1\overline{\sigma }}^{\dag }d_{1\overline{\sigma }%
}\right\rangle =\left\langle d_{1\overline{\sigma }}^{\dag }c_{\mathbf{k}1%
\overline{\sigma }}\right\rangle $, the higher-order Green
function is thus derived as
\begin{eqnarray}
&&\left\langle \left\langle n_{1\overline{\sigma }}d_{1\sigma
}|d_{1\sigma
}^{\dagger }\right\rangle \right\rangle   \notag\\
&&\simeq \frac{\left\langle n_{1%
\overline{\sigma }}\right\rangle }{i\omega _{n}-\epsilon
_{1}-U}\left[ 1-t\left\langle \left\langle d_{2\sigma }|d_{1\sigma
}^{\dagger
}\right\rangle \right\rangle \right.  \notag \\
&& +\left( \sum_{\mathbf{k}}\frac{\left| V_{1}\right|
^{2}}{i\omega
_{n}-\epsilon _{\mathbf{k}1}}+\sum_{\mathbf{k}}\frac{\left| V_{2}\right| ^{2}%
}{i\omega _{n}-\epsilon _{\mathbf{k}2}}\right)  \notag\\
&& \left. \times \left\langle \left\langle d_{1\sigma }|d_{1\sigma
}^{\dagger }\right\rangle \right\rangle \right].
\end{eqnarray}%
Introducing
\begin{eqnarray}
F_{1}\left( i\omega _{n},\epsilon _{1},\left\langle n_{1\overline{\sigma }%
}\right\rangle \right) &=&1+\frac{U\left\langle n_{1\overline{\sigma }%
}\right\rangle }{i\omega _{n}-\epsilon _{1}-U},  \notag \\
F_{2}\left( i\omega _{n},\epsilon _{2},\left\langle n_{2\overline{\sigma }%
}\right\rangle \right) &=&1+\frac{U\left\langle n_{2\overline{\sigma }%
}\right\rangle }{i\omega _{n}-\epsilon _{2}-U},
\end{eqnarray}
the dot Green function $\left\langle \left\langle d_{1\sigma
}|d_{1\sigma }^{\dagger }\right\rangle \right\rangle $ is found to
satisfy the following equation:
\begin{eqnarray}
&&\left[ i\omega _{n}-\epsilon _{1}-F_{1}\left( \sum_{\mathbf{k}}\frac{%
\left| V_{1}\right| ^{2}}{i\omega _{n}-\epsilon _{\mathbf{k}1}}+\sum_{%
\mathbf{k}}\frac{\left| V_{2}\right| ^{2}}{i\omega _{n}-\epsilon _{\mathbf{k}%
2}}\right) \right]  \notag \\
&&\times \left\langle \left\langle d_{1\sigma }|d_{1\sigma
}^{\dagger }\right\rangle \right\rangle +tF_{1}\left\langle
\left\langle d_{2\sigma }|d_{1\sigma }^{\dagger }\right\rangle
\right\rangle =F_{1}.
\end{eqnarray}
By applying the similar method, the equation of motion of the
Green function $\left\langle \left\langle d_{2\sigma }|d_{1\sigma
}^{\dagger }\right\rangle \right\rangle $ satisfies:
\begin{eqnarray}
&&tF_{2}\left\langle \left\langle d_{1\sigma }|d_{1\sigma
}^{\dagger
}\right\rangle \right\rangle  \notag \\
&&+\left[ i\omega _{n}-\epsilon _{2}-F_{2}\left( \sum_{\mathbf{k}}\frac{%
\left| V_{3}\right| ^{2}}{i\omega _{n}-\epsilon _{\mathbf{k}3}}+\sum_{%
\mathbf{k}}\frac{\left| V_{4}\right| ^{2}}{i\omega _{n}-\epsilon _{\mathbf{k}%
4}}\right) \right]  \notag \\
&&\times \left\langle \left\langle d_{2\sigma }|d_{1\sigma
}^{\dagger }\right\rangle \right\rangle =0.
\end{eqnarray}
From the above two equations, we obtain the retarded Green
functions as shown in Eqs. (\ref{GF11}) and (\ref{GF12}).
Similarly, we can obtain the retarded Green functions
$\left\langle \left\langle d_{2\sigma }|d_{2\sigma }^{\dagger
}\right\rangle \right\rangle$ and $\left\langle \left\langle
d_{1\sigma }|d_{2\sigma }^{\dagger }\right\rangle \right\rangle$
as shown in Eqs. (\ref{GF22}) and (\ref{GF12}).

%\appendix
\section{The Evaluation of the Keldysh Less Green Functions}

With the help of equation-of-motion approach, one can also obtain
the Keldysh less Green functions, which is needed in the
self-consistent evaluation of the dot occupation numbers when a
finite bias voltage is applied across the system. Its
equation of motion is yielded as%
\begin{eqnarray}
&&i\frac{\partial }{\partial t}G_{1\sigma ,1\sigma }^{<}\left( t,t^{\prime
}\right)  \notag \\
&&=\epsilon _{1}i\left\langle d_{1\sigma }^{\dagger }\left( t^{\prime
}\right) d_{1\sigma }\left( t\right) \right\rangle -ti\left\langle
d_{1\sigma }^{\dagger }\left( t^{\prime }\right) d_{2\sigma }\left( t\right)
\right\rangle  \notag \\
&&+Ui\left\langle d_{1\sigma }^{\dagger }\left( t^{\prime }\right) \left[
n_{1\overline{\sigma }}d_{1\sigma }\right] \left( t\right) \right\rangle
+V_{1}^{\ast }\sum_{\mathbf{k}}i\left\langle d_{1\sigma }^{\dagger }\left(
t^{\prime }\right) c_{\mathbf{k}1\sigma }\left( t\right) \right\rangle
\notag \\
&&+V_{2}^{\ast }\sum_{\mathbf{k}}i\left\langle d_{1\sigma }^{\dagger }\left(
t^{\prime }\right) c_{\mathbf{k}2\sigma }\left( t\right) \right\rangle ,
\end{eqnarray}%
and then we perform the Fourier transform%
\begin{eqnarray}
&&\omega G_{1\sigma ,1\sigma }^{<}\left( \omega \right)  \notag \\
&&=\epsilon _{1}G_{1\sigma ,1\sigma }^{<}\left( \omega \right) -tG_{2\sigma
,1\sigma }^{<}\left( \omega \right) +V_{2}^{\ast }\sum_{\mathbf{k}}G_{%
\mathbf{k}2\sigma ,1\sigma }^{<}\left( \omega \right)  \notag \\
&&+U\left\langle \left\langle n_{1\overline{\sigma }}d_{1\sigma }|d_{1\sigma
}^{\dagger }\right\rangle \right\rangle ^{<}\left( \omega \right)
+V_{1}^{\ast }\sum_{\mathbf{k}}G_{\mathbf{k}1\sigma ,1\sigma }^{<}\left(
\omega \right) .
\end{eqnarray}%
Under the same Hartree-Fock approximation as in the retarded Green
functions, the equation of motion for the
higher-order Keldysh Green function $\left\langle \left\langle n_{1\overline{%
\sigma }}d_{1\sigma }|d_{1\sigma }^{\dagger }\right\rangle \right\rangle
^{<} $ can also be deduced to%
\begin{eqnarray}
&&\left( \omega -\epsilon _{1}-U\right) \left\langle \left\langle n_{1%
\overline{\sigma }}d_{1\sigma }|d_{1\sigma }^{\dagger }\right\rangle
\right\rangle ^{<}\left( \omega \right)  \notag \\
&=&-t\left\langle n_{1\overline{\sigma }}\right\rangle G_{2\sigma ,1\sigma
}^{<}\left( \omega \right) +V_{1}^{\ast }\left\langle n_{1\overline{\sigma }%
}\right\rangle \sum_{\mathbf{k}}G_{\mathbf{k}1\sigma ,1\sigma }^{<}\left(
\omega \right)  \notag \\
&&+V_{2}^{\ast }\left\langle n_{1\overline{\sigma }}\right\rangle \sum_{%
\mathbf{k}}G_{\mathbf{k}2\sigma ,1\sigma }^{<}\left( \omega \right) .
\end{eqnarray}%
By using the exact Dyson equation for $G_{\mathbf{k}1(2)\sigma ,1\sigma }^{<}
\left( \omega \right)$:%
\begin{eqnarray}
G_{\mathbf{k}1(2)\sigma ,1\sigma }^{<}\left( \omega \right) &=&V_{1}\left[
g_{\mathbf{k}1\left( 2\right) \sigma }^{r}\left( \omega \right) G_{1\sigma
,1\sigma }^{<}\left( \omega \right) \right.  \notag \\
&&\left. +g_{\mathbf{k}1\left( 2\right) \sigma }^{<}\left( \omega \right)
G_{1\sigma ,1\sigma }^{a}\left( \omega \right) \right] ,
\end{eqnarray}%
the equation of motion for the $G_{1\sigma ,1\sigma }^{<}$ becomes
\begin{widetext}
\begin{equation}
\left[ \omega -\epsilon _{1}-F_{1} \left( \left\vert
V_{1}\right\vert ^{2}\sum_{\mathbf{k}}g_{\mathbf{k}1\sigma
}^{r}+\left\vert
V_{2}\right\vert ^{2}\sum_{\mathbf{k}}g_{\mathbf{k}2\sigma }^{r}\right) %
\right] G_{1\sigma ,1\sigma }^{<}+tF_{1} G_{2\sigma ,1\sigma }^{<}
=F_{1} \left[ \left\vert V_{1}\right\vert ^{2}\sum_{%
\mathbf{k}}g_{\mathbf{k}1\sigma }^{<}+\left\vert V_{2}\right\vert ^{2}\sum_{%
\mathbf{k}}g_{\mathbf{k}2\sigma }^{<}\right] G_{1\sigma ,1\sigma }^{a}.
\end{equation}
By applying the same method, we obtain the equation of motion for the Keldysh less Green function $%
G_{2\sigma ,1\sigma }^{<}$ as
\begin{equation}
tF_{2} G_{1\sigma ,1\sigma }^{<}+\left[ \omega -\epsilon
_{2}-F_{2} \left( \left\vert V_{3}\right\vert
^{2}\sum_{\mathbf{k}}g_{\mathbf{k}3\sigma }^{r}+\left\vert
V_{4}\right\vert ^{2}\sum_{\mathbf{k}}g_{\mathbf{k}4\sigma }^{r}\right) %
\right] G_{2\sigma ,1\sigma }^{<}
=F_{2} \left[ \left\vert V_{3}\right\vert ^{2}\sum_{%
\mathbf{k}}g_{\mathbf{k}3\sigma }^{<}+\left\vert V_{4}\right\vert ^{2}\sum_{%
\mathbf{k}}g_{\mathbf{k}4\sigma }^{<}\right] G_{2\sigma ,1\sigma }^{a}.
\end{equation}%
\end{widetext}
From these two equations, we can solve the Keldysh less Green
functions as Eq. (\ref{G_less_11}) for dot 1 and Eq.
(\ref{G_less_22}) for dot 2.

\end{document}